\documentclass[a4]{article}
\usepackage{hyperref}
\usepackage{graphicx,epsfig}
\input{talk.mac}

\title{New phenomenology from an old theory--the BCS theory of superconductivity revisited}
\author{Drago\c s-Victor Anghel\thanks{Institutul National de C\&D pentru Fizica si Inginerie Nucleara -- Horia Hulubei, {\tt dragos@theory.nipne.ro}}}

\date{\today}

\begin{document}

\maketitle
% \pacs{05.30.-d}{Quantum statistical mechanics}
% \pacs{05.30.Ch}{Quantum ensemble theory}
% \pacs{74.20.Fg}{BCS theory and its development}

\begin{abstract}
I analyze the low temperature limit of the BCS theory of s-wave single-band superconductors, when the attraction band may be asymmetric with respect to the chemical potential.
I discuss equilibrium systems, taking consistently into account the variation of the energy and of the total number of particles with the populations of the quasiparticle energy levels.
I show that the equation for the energy gap has two solutions, one of which is stable and the other one is metastable.
When the chemical potential is the center of the attraction band (the standard BCS assumption), the energy gap in the stable solution is $\Delta_0$, whereas in the metastable one is $\Delta_0/3$.
If the chemical potential is not in the center of the attraction band, then a quasiparticle imbalance appears.
If the absolute value of the difference between the chemical potential and center of the attraction band is bigger than $2\Delta_0$, then the superconducting energy gap cannot be formed.
If the number of particles is conserved and the attraction band is asymmetric, then the stable solution is unphysical and only metastable solutions are realized.
\end{abstract}

\maketitle

\section{Introduction} \label{intro}

In Ref. \cite{PhysicaA.464.74.2016.Anghel}, the Bardeen-Cooper-Schrieffer (BCS) theory of superconductivity \cite{PhysRev.108.1175.1957.Bardeen,Tinkham:book} has been revisited under the assumption that the attraction band--the single-particle energy interval in which the pairing interaction is manifested--is asymmetric with respect to the chemical potential of the system.
Surprisingly, this asymmetry changes dramatically the phenomenology of the superconducting phase: a quasiparticle imbalance appears in equilibrium, the energy gap changes, as well as the critical temperature.
Furthermore, in grandcanonical conditions, the normal metal-superconductor phase transition may become discontinuous (see for example \cite{NuclPhysA.887.1.2012.Parvan} for a discussion related to the order of phase transitions in different ensembles).
The formalism is also applicable to nuclear matter.

Asymmetric attraction bands with respect to the chemical potential are known in the context of multi-band superconductors or whenever the Fermi energy is close to one extremum of the conduction band (see for example \cite{LowTempPhys.41.112.2015.Parfeniev, PhysRevLett.87.047001.2001.Bouquet, PhysRevLett.87.137005.2001.Szabo, Science.314.1910.2006.Tanaka, PhysRevLett.98.267004.2007.Kondo}).
In such cases, the asymmetry is imposed by the limits of the conduction band.

If the attraction band is denoted by  $I_V \equiv [\mu - \hbar\omega_c, \mu + \hbar\omega_c]$ and the chemical potential is denoted by $\mu_R$, then the standard BCS phenomenology is recovered only if $\mu = \mu_R$.
Otherwise, the energy gap and the populations of the quasiparticle energy levels are calculated by solving a system of integral equations.
In Ref.~\cite{PhysicaA.464.74.2016.Anghel} the system was analyzed in the grandcanonical ensemble and it was shown that the phase transition temperature decreases with $|\mu_R - \mu|$.
Furthermore, the system of equations for the energy gap may have more than one solution at fixed temperature and chemical potential.

Quasiparticle imbalance in the context of the BCS theory have been reported before for non-equilibrium superconductors (see for example Refs. \cite{PhysRevLett.28.1363.1972.Clarke, PhysRevLett.28.1366.1972.Tinkham, PhysRevB.6.1747.1972.Tinkham, PhysRevB.22.4346.1980.Smith, PhysRevB.21.3879.1980.Smith}).
Such non-equilibrium situations can be described also by our approach, but here I focus only on equilibrium superconductivity.

Equilibrium quasiparticle imbalance \cite{PhysRevLett.72.558.1994.Hirsch, PhysRevB.58.8727.1998.Hirsch, PhysScr.88.035704.2013.Hirsch} appears also in the model of hole superconductivity \cite{PhysRevB.39.11515.1989.Hirsch, PhysRevB.41.6435.1990.Marsiglio}, but the concept and the predictions of this model are, in many respects, very different from the BCS theory.
I do not make here comparisons between different models.

% The paper is organized as follows. In the next section I present the basic equations, in the third section I present the low temperature limit of these equations and their solutions, whereas in the forth section I present the conclusions.

In this paper I analyze the energy gap and the populations in the low temperature limit in order to determine the number of solutions and their stability in the grandcanonical ensemble.
% I show that the system has two solutions if $|\mu_R-\mu| < 2\Delta_0$ and no solutions otherwise.
% I impose conservation of the (average) number of particles
Afterwards, I impose the conservation of the (average) number of particles, in order to obtain the results corresponding to the canonical ensemble.

\section{The formalism} \label{sec_formalism}

In Ref. \cite{PhysicaA.464.74.2016.Anghel}, following the standard procedure (see for example Ref. \cite{Tinkham:book}) and maximizing of the grandcanonical partition function of a superconductor, the populations of the quasiparticle states were obtained in the form
\begin{subequations} \label{eqs_pop}
\begin{equation}
  n_{\bk i} = \frac{1}{e^{\beta(\epsilon_{\bk}-\tilde\mu)}+1}, \quad i=0,1 , \label{pop_til_eps}
\end{equation}
where $i$ indicates the type of quasiparticle,
\begin{equation}
  \tilde\mu \equiv \frac{\mu_R - \mu}{\epsilon_\bk} \left[ \xi_\bk - \frac{ \sum_\bk \left( 1 - n_{\bk 0} - n_{\bk 1} \right) \xi_\bk \epsilon_\bk^{-3}}
  { \sum_\bk \left(1 - n_{\bk 0} - n_{\bk 1} \right) \epsilon_\bk^{-3} } \right] . \label{def_tilde_mu} \\
  %
%   \tilde\mu \equiv \frac{\mu_R - \mu}{\epsilon_\bk} \left[ \xi_\bk - \frac{ \int_{-\hbar\omega_c}^{\hbar\omega_c} \sigma(\xi+\mu) ( 1 - n_{\xi 0} - n_{\xi 1} ) \frac{\xi}{\epsilon^3} \, d\xi }
%   { \int_{-\hbar\omega_c}^{\hbar\omega_c} \frac{(1 - n_{0\xi} - n_{1\xi}) \sigma(\xi+\mu) d\xi}{\epsilon^3} } \right] . \label{def_tilde_mu}
\end{equation}
\end{subequations}
is an ``effective'' chemical potential, $\bk$ is the wavevector, $\xi_\bk \equiv \epsilon_\bk^{(0)} - \mu$ is the difference between the electron's free particle energy and the center of the attraction band, $\beta \equiv 1/(k_BT)$ is the inverse temperature, and $\epsilon_\bk \equiv \sqrt{\xi_\bk^2 + \Delta^2}$ is the BCS quasiparticle energy--both, $\tilde \mu$ and $\epsilon_\bk$ are independent of $i$.
The BCS energy gap $\Delta$ is determined from the equation
\begin{equation}
  1 = \frac{V}{2} \sum_\bk \frac{1 - n_{\bk 0} - n_{\bk 1}}{\epsilon_\bk} , \label{def_Delta2}
\end{equation}
where $V$ is the pairing potential, independent of the pairs momenta and different from zero if and only if the single-particle energies of the electrons forming the pairs are within the attraction band $I_V$ (standard BCS assumption).
Equations (\ref{eqs_pop}) and (\ref{def_Delta2}) should be solved self-consistently to obtain the populations and the energy gap of the superconductor.

In the quasicontinuous limit, the summation over $\bk$ is transformed into an integral over $\xi$.
If the density of states  (DOS) is constant, $\sigma(\xi) \equiv \sigma_0$, then Eq. (\ref{def_Delta2}) simplifies to
\begin{eqnarray}
  \frac{2}{\sigma_0V} &=& \int_{-\hbar\omega_c}^{\hbar\omega_c} \frac{1 - n_{\xi 0} - n_{\xi 1}}{\sqrt{\xi^2+\Delta^2}} d\xi .
  \label{Eq_int_Delta1}
\end{eqnarray}
In the zero temperature limit, if we set $n_{\xi 0} = n_{\xi 1} = 0$ for any $\xi$ (we shall see further that there is a metastable state in which this condition is not satisfied), one can obtain an analytical expression for the energy gap in the weak coupling limit ($\sigma_0V \ll 1$): $\Delta_0 = 2\hbar\omega_c \exp[-1/(\sigma_0 V)]$.
Similarly, at the critical temperature $T_c$ the energy gap should be zero, and form Eq. (\ref{Eq_int_Delta1}) one obtains $k_B T_c = A \hbar\omega_c e^{-1/(\sigma_0 V)}$, where $A = 2 e^\gamma/\pi \approx 1.13$ and $\gamma\approx 0.577$ is the Euler's constant \cite{Tinkham:book}.

To be able to solve the self-consistent set of equations (\ref{eqs_pop}) and (\ref{def_Delta2}), we first identify in Eq. (\ref{def_tilde_mu}) the constant $F$, such that the effective chemical potential is written $\tilde \mu \equiv (\mu_R-\mu)(\xi-F)/\epsilon$.
% %
% \begin{equation}
%   F \equiv \frac{ \int_{-\hbar\omega_c}^{\hbar\omega_c} \sigma(\xi+\mu) ( 1 - n_{\xi 0} - n_{\xi 1} ) \frac{\xi}{\epsilon^3} \, d\xi }
%   { \int_{-\hbar\omega_c}^{\hbar\omega_c} \frac{(1 - n_{0\xi} - n_{1\xi}) \sigma(\xi+\mu) d\xi}{\epsilon^3} }
%   \label{def_F0}
% \end{equation}
%
If the DOS is constant, Eqs. (\ref{eqs_pop}) get a simpler form and are equivalent to
\begin{subequations}\label{def_F_sigma0_set}
\begin{eqnarray}
  F &\equiv& \frac{ \int_{-\hbar\omega_c}^{\hbar\omega_c} ( 1 - n_{\xi 0} - n_{\xi 1} ) \frac{\xi}{\epsilon^3} \, d\xi }
  { \int_{-\hbar\omega_c}^{\hbar\omega_c} \frac{(1 - n_{\xi 0} - n_{\xi 1}) d\xi}{\epsilon^3} } , \label{def_F_sigma0} \\
  n_{\xi i} &=& \frac{1}{e^{\beta[\epsilon_{\xi}-(\mu_R-\mu)(\xi - F)/\epsilon_\xi]}+1} . \label{pop_til_eps_sigma0} 
\end{eqnarray}
\end{subequations}
Introducing the dimensionless variables $x_F \equiv \beta F$, $x\equiv \beta\epsilon$, $y \equiv \beta\Delta$, and $y_R \equiv \beta (\mu_R - \mu)$, and assuming a constant density of states, the set of equations (\ref{Eq_int_Delta1}) and (\ref{def_F_sigma0_set}) can be transformed into
\begin{subequations}\label{def_xF_sigma0_set}
\begin{eqnarray}
  x_F &=& \frac{\int_{y}^{\beta\hbar\omega_c} \frac{( n_{-\xi_x} - n_{\xi_x} )\, dx}{x^2}}{\int_{y}^{\beta\hbar\omega_c} \frac{(1 - n_{-\xi_x} - n_{\xi_x})\, dx}{x^2 \sqrt{x^2-y^2}}} , \label{def_xF_sigma0} \\
  n_{\xi_x} &=& \frac{1}{e^{x-y_R\left(\sqrt{x^2-y^2} - x_F\right)/x}+1} , \label{pop_til_eps_sigma0_p} \\
  n_{-\xi_x} &=& \frac{1}{e^{x-y_R\left(-\sqrt{x^2-y^2} - x_F\right)/x}+1} \label{pop_til_eps_sigma0_n} \\
  \frac{1}{\sigma_0 V} &=& \int_y^{\beta\hbar\omega_c} \frac{1 - n_{-\xi_x} - n_{\xi_x}}{\sqrt{x^2 - y^2}} dx % \equiv I_\Delta (y)
  \label{Eq_int_Delta1_2}
\end{eqnarray}
\end{subequations}
where we wrote explicitly the populations for the positive and negative branches, namely $\xi = \sqrt{\epsilon^2 - \Delta^2}$ in Eq. (\ref{pop_til_eps_sigma0_p}) and $\xi = - \sqrt{\epsilon^2 - \Delta^2}$ in Eq. (\ref{pop_til_eps_sigma0_n}).
We observe that Eqs. % (\ref{def_F_sigma0_set}) and
(\ref{def_xF_sigma0_set}) are symmetric under the exchange $y_R \to -y_R$, $x_F \to -x_F$, and $\xi \to -\xi$.
Solving self-consistently the set of equations (\ref{def_xF_sigma0_set}) we obtain the equilibrium populations and $\Delta$.
% The total number of particles is \cite{PhysicaA.464.74.2016.Anghel}
% %
% \begin{equation}
%   N = N_\mu + 2 \sigma_0 \int_{-\hbar\omega_c}^{\hbar\omega_c} \frac{\xi n_\xi}{\sqrt{\xi^2 + \Delta^2}} d\xi , \label{N_tot}
% \end{equation}
% %
% where $N_\mu$ (a constant) represents the number of single-particle states up to level $\mu$, in the noninteracting system.

The system (\ref{def_xF_sigma0_set}) depends on two parameters: $y_R$ and $\beta = 1/(k_BT)$.
In the next section we shall study the solutions of the system in the limit $\beta \to \infty$ (or $T\to 0$) for different values of the parameter $\mu_R - \mu = y_R/\beta$.

\section{Low temperature limit and constant DOS} \label{sec_lowT}

% I start from the system of Eqs. (\ref{def_xF_sigma0_set}) and we discuss only the case $y_R > 0$.
Since the solutions for $y_R < 0$ can be obtained from the solutions with $y_R>0$, by the replacement $x_F \to - x_F$ and exchanging $n_\xi$ with $n_{-\xi}$, in the following we shall study the system (\ref{def_xF_sigma0_set}) only in the case $y_R \ge 0$.
For this, we analyze the argument of the exponential function in the denominator of $n_{\xi_x}$ and $n_{-\xi_x}$.
If we write $n_{\xi_x} \equiv \left\{ \exp[\beta m_{\xi_x}] + 1 \right\}^{-1}$ and $n_{-\xi_x} \equiv \left\{ \exp[\beta m_{-\xi_x}] + 1 \right\}^{-1}$, then
\begin{subequations} \label{defs_mx}
\begin{eqnarray}
  m_{\xi_x}    &\equiv& \frac{\Delta}{r} \left( r^2 - a\sqrt{r^2 - 1} + ab \right) , \label{def_mx} \\
  m_{-\xi_x} &\equiv& \frac{\Delta}{r} \left( r^2 + a \sqrt{r^2 - 1} + ab \right) , \label{def_mmx}
\end{eqnarray}
\end{subequations}
where $r = \epsilon/\Delta = x/y \ge 1$, $a = (\mu_R - \mu)/\Delta = y_R/y$, and $b = F/\Delta = x_F/y$.
When $m_{\xi_x}>0$, then $\lim_{T\to0}\beta m_{\xi_x} = \infty$ and $\lim_{T\to0} n_{\xi_x} = 0$, whereas if $m_{\xi_x}<0$, then $\lim_{T\to0}\beta m_{\xi_x} = -\infty$ and $\lim_{T\to0} n_{\xi_x} = 1$.
Similarly, in the limit $T\to0$, if $m_{-\xi_x}>0$, then $n_{-\xi_x}=0$ and if $m_{-\xi_x}<0$, then $n_{-\xi_x}=1$.

Let us now find the values of $r$ for which $m_{\xi_x} < 0$ or $m_{-\xi_x} < 0$. In Eqs. (\ref{defs_mx}) we denote $t \equiv \sqrt{r^2 - 1} \ge 0$ and we rewrite them as
\begin{subequations} \label{defs_mx3}
\begin{eqnarray}
  m_{\xi_x}    &\equiv& \frac{\Delta}{\sqrt{t^2+1}} ( t^2 - at + ab + 1 ) , \label{def_mx3} \\
  m_{-\xi_x} &\equiv& \frac{\Delta}{\sqrt{t^2+1}} ( t^2 + at + ab + 1 ) , \label{def_mmx3}
\end{eqnarray}
\end{subequations}
We can now see that $m_{\xi_x}$ and $m_{-\xi_x}$ may take negative values only if the discriminant of Eqs. (\ref{defs_mx3}), $D \equiv a^2 - 4ab - 4$, is positive.
Then, the solutions for Eq. (\ref{def_mx3})  are
\begin{equation}
  t_1 = \frac{a - \sqrt{a^2 - 4ab - 4}}{2} \ {\rm and} \
  t_2 = \frac{a + \sqrt{a^2 - 4ab - 4}}{2} ,
  \label{t12}
\end{equation}
whereas the solutions for Eq. (\ref{def_mmx3})  are $t'_1 = - t_2$ and $t'_2 = - t_1$.
Obviously, $t_2 > 0$ and $t_1' < 0$.
If we denote by $I_r \equiv (r_1,r_2)$, the interval on which $m_{\xi_x}(r) < 0$ (\ref{def_mx}), then
% Equation (\ref{t12}) leads to $r_{1,2}^2 = t_{1,2}^2 + 1$, namely
%
\begin{subequations}\label{Eqs_mr3}
\begin{equation}
  r_2 \equiv \sqrt{t_2^2 + 1} = \sqrt{\frac{a}{2}\left( a-2b + \sqrt{ a^2 - 4ab - 4 } \right)} \ge 1 . \label{Eq_mr3}
\end{equation}
Since $t_1 \le 0$ if and only if $ab \le -1$, then
\begin{equation}
  r_1 = \left\{ \begin{array}{l}
                    \sqrt{\frac{a}{2}\left( a-2b - \sqrt{ a^2 - 4ab - 4 } \right)} ,\ {\rm if}\ ab > -1 , \\
                    1,\ {\rm if}\ ab \le -1 .
                  \end{array} \right. \label{Eq_mr4}
\end{equation}
Similarly, $I_r' \equiv (r_1',r_2')$ is the interval on which $m_{-\xi_x}(r) < 0$ (\ref{def_mmx}). Then $r_1' = 1$ (since $t_1' = -t_2 < 0$) and
\begin{equation}
  r_2' = \left\{
  \begin{array}{l}
    \sqrt{\frac{a}{2} \left(a-2b - \sqrt{ a^2 - 4ab - 4 } \right)} ,\ {\rm if}\ ab < -1 , \\
    1,\ {\rm if}\ ab \ge -1 .
  \end{array} \right. \label{Eq_mr5}
\end{equation}
\end{subequations}
For $r \in I_r$, $\lim_{T\to0}n_{\xi_x}(T) = 1$, whereas for $r \in [1,\infty) \setminus [r_1,r_2]$, $\lim_{T\to0}n_{\xi_x}(T) = 0$. Similarly, for $r \in I_r'$, $\lim_{T\to0}n_{-\xi_x}(T) = 1$, whereas for $r \in [1,\infty) \setminus [r'_1,r'_2]$, $\lim_{T\to0}n_{-\xi_x}(T) = 0$.
We also observe that $I_r' \subset I_r$, because, if $t_1 \ge 0$, then $r_1 \ge 1$ and $I_r' = \emptyset \subset I_r$, whereas if $t_1 \le 0$, then $r_1 = r_1' = 1$ and $r_2 > r_2'$ (Eqs. \ref{Eq_mr3} and \ref{Eq_mr5}), which implies again $I_r' \subset I_r$.
Using this observation we see that $f_n(x) \equiv n_{-\xi_x} - n_{\xi_x}$, which appears in the integrand in the numerator of Eq. (\ref{def_xF_sigma0}), is different from zero only if $x/y = r \in {\rm Int}( I_r \setminus I_r')$ (where ${\rm Int}(\cdot)$ denotes the interior of an interval), whereas $f_d(x) \equiv 1 - n_{-\xi_x} - n_{\xi_x}$, which appears in the integrand in the denominator, is zero in the same interval. Furthermore, $f_d(x) = -1$, if $x/y = r \in I_r'$, and $f_d(x) = 1$, if $x/y = r \in (r_2, \infty)$.

Let us now calculate $\Delta$ and $x_F$. I introduce the notation $r_0 \equiv \sqrt{(a/2) \left(a-2b - \sqrt{ a^2 - 4ab - 4 } \right)} \ge 1$.
If $ab \ge -1$, then, from Eqs. (\ref{def_xF_sigma0_set}), we obtain
\begin{subequations} \label{Eqs_Delta_xF_T0}
\begin{eqnarray}
  \frac{1}{\sigma_0 V} % &=& \int_y^{r_0 y} \frac{dx}{\sqrt{x^2 - y^2}} + \int_{r_2 y}^{\sqrt{(\beta\hbar\omega_c)^2 + y^2}} \frac{dx}{\sqrt{x^2 - y^2}} \nonumber \\
  &=& \log\left( \frac{2 \hbar\omega_c}{\Delta} \right) - \log \left( \frac{r_2 + \sqrt{r_2^2 - 1}}{r_0 + \sqrt{r_0^2 - 1}} \right) , \label{Eqs_Delta_T0} \\ 
  b % &=& \frac{- \int_{r_0}^{r_2} \frac{dr}{r^2}}{\int_{1}^{r_0} \frac{dr}{r^2 \sqrt{r^2-1}} + \int_{r_2}^{\beta\hbar\omega_c/y} \frac{dr}{r^2 \sqrt{r^2-1}}} \nonumber \\
  &=& \frac{\frac{1}{r_2} - \frac{1}{r_0}}{1 - \frac{\sqrt{r_2^2 - 1}}{r_2} + \frac{\sqrt{r_0^2 - 1}}{r_0}} , \label{Eqs_xF_T0}
\end{eqnarray}
and we observe that $b < 0$.
From Eq. (\ref{Eqs_Delta_T0}) we can eliminate $1/(\sigma_0 V)$ and write
\begin{equation}
%   \log\left( \frac{\Delta_0}{\Delta} \right) = \log \left( \frac{r_2 + \sqrt{r_2^2 - 1}}{r_0 + \sqrt{r_0^2 - 1}} \right) , \label{Eqs_Delta_T0_1}
  \frac{\Delta}{\Delta_0} = \frac{r_0 + \sqrt{r_0^2 - 1}}{r_2 + \sqrt{r_2^2 - 1}} , \label{Eqs_Delta_T0_1}
\end{equation}
\end{subequations}
If $ab < -1$, then, 
\begin{subequations} \label{Eqs_Delta_xF_T0_v2}
\begin{eqnarray}
  \frac{1}{\sigma_0 V} % &=& - \int_y^{r_0 y} \frac{dx}{\sqrt{x^2 - y^2}} + \int_{r_2 y}^{\sqrt{(\beta\hbar\omega_c)^2 + y^2}} \frac{dx}{\sqrt{x^2 - y^2}} \nonumber \\
  &=& \log\left( \frac{2 \hbar\omega_c}{\Delta} \right) - \log \left( r_2 + \sqrt{r_2^2 - 1} \right) \nonumber \\
  && - \log \left( r_0 + \sqrt{r_0^2 - 1} \right) , \label{Eqs_Delta_T0_v2} \\ 
  b % &=& \frac{- \int_{r_0}^{r_2} \frac{dr}{r^2}}{- \int_{1}^{r_0} \frac{dr}{r^2 \sqrt{r^2-1}} + \int_{r_2}^{\beta\hbar\omega_c/y} \frac{dr}{r^2 \sqrt{r^2-1}}} \nonumber \\
  &=& \frac{\frac{1}{r_2} - \frac{1}{r_0}}{1 - \frac{\sqrt{r_2^2 - 1}}{r_2} - \frac{\sqrt{r_0^2 - 1}}{r_0}} , \label{Eqs_xF_T0_v2}
\end{eqnarray}
Eliminating $1/(\sigma_0 V)$ from Eq. (\ref{Eqs_Delta_T0_v2}) we write
\begin{equation}
%   \log\left( \frac{\Delta_0}{\Delta} \right) = \log \left( \frac{r_2 + \sqrt{r_2^2 - 1}}{r_0 + \sqrt{r_0^2 - 1}} \right) , \label{Eqs_Delta_T0_1}
  \frac{\Delta}{\Delta_0} = \frac{1}{\left( r_0 + \sqrt{r_0^2 - 1} \right) \left(r_2 + \sqrt{r_2^2 - 1} \right)} . \label{Eqs_Delta_T0_1_v2}
\end{equation}
\end{subequations}
If $ab = -1$, then $r_0 = 1$ and Eqs. (\ref{Eqs_Delta_xF_T0}) and (\ref{Eqs_Delta_xF_T0_v2}) give the same results, implying that the functions $b(a)$ and $\Delta(a)$ are continuous.

\begin{figure}[t]
  \centering
  \includegraphics[width=8cm,bb=0 0 694 572,keepaspectratio=true]{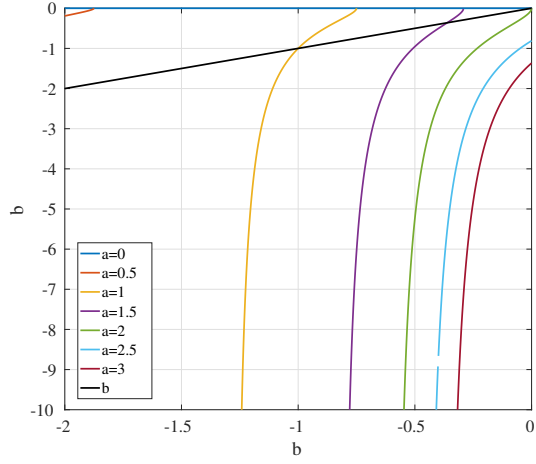}
  % function_b.eps: 0x0 pixel, 300dpi, 0.00x0.00 cm, bb=0 0 694 572
  \caption{(Color online) The r.h.s. of Eq. (\ref{Eqs_xF_T0}), for $-1 \le ab \le 0$, continued by (\ref{Eqs_xF_T0_v2}), for $ab \le -1$, plotted vs. $b$, for different values of $a$. The straight black line is $b$ vs $b$.}
  \label{function_b}
\end{figure}

Numerical and analytical analysis of Eqs. (\ref{Eqs_Delta_xF_T0}) and (\ref{Eqs_Delta_xF_T0_v2}) show that for $0 < a < 2$ there are two solutions (see Fig.~\ref{function_b}): one with $b = 0$, $n_{\xi_x}(T=0) = n_{-\xi_x}(T=0) = 0$, and $\Delta(T=0) = \Delta_0$, and another one, with $b < 0$ and $\Delta(T=0) < \Delta_0$, whereas $n_{\xi_x}(T=0)$ and $n_{-\xi_x}(T=0)$ may take nonzero values for some values of $x$. For $a = 2$ (green curve in Fig. \ref{function_b}), only the solution with $b = 0$ remains, whereas for $a > 2$, Eqs. (\ref{Eqs_Delta_xF_T0}) and (\ref{Eqs_Delta_xF_T0_v2}) have no solutions and the superconducting phase cannot be formed at $T=0$.

\begin{figure}[t]
  \centering
  \includegraphics[width=70mm,bb=0 0 694 575,keepaspectratio=true]{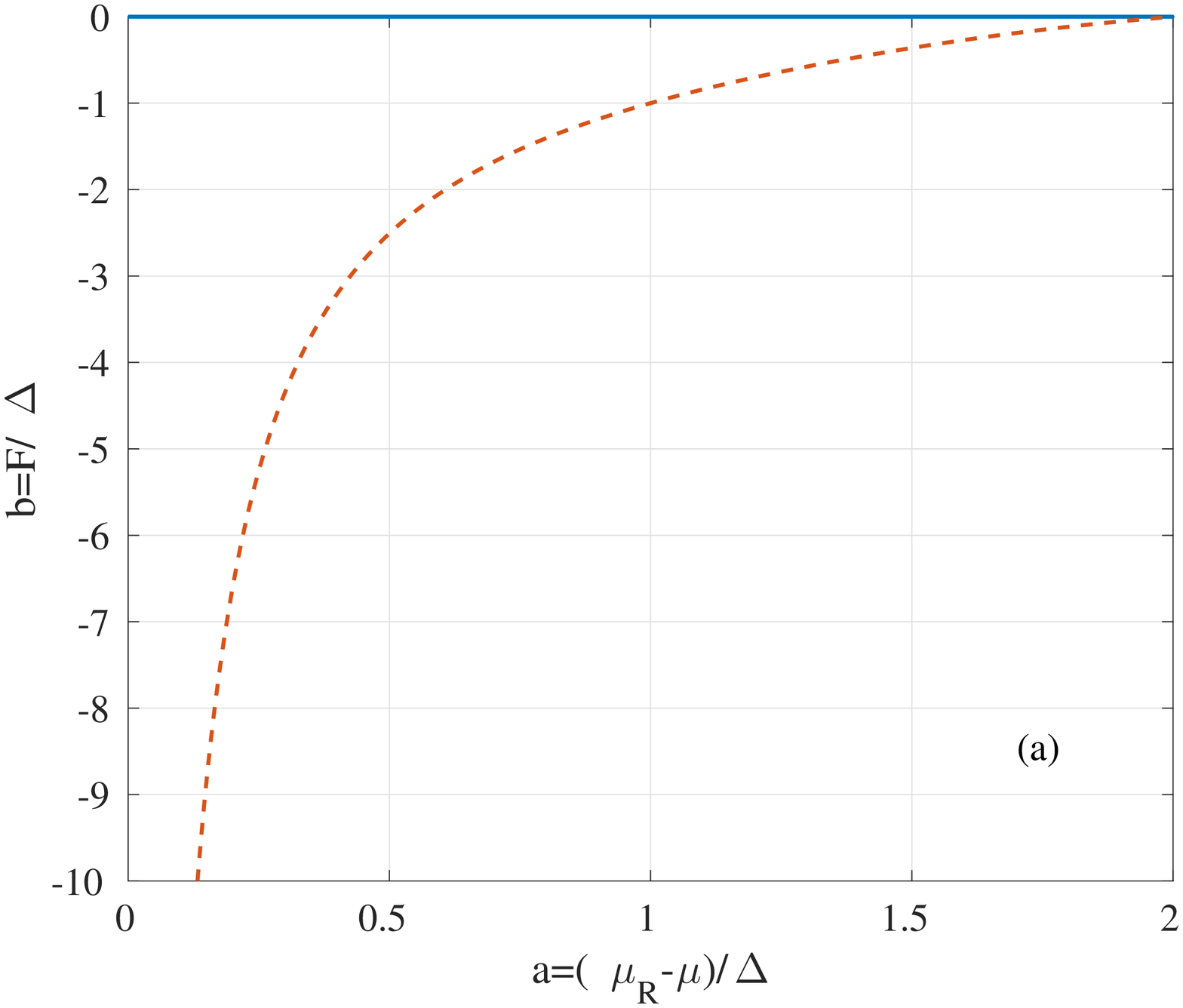}
  % Delta_ratio_T0.eps: 0x0 pixel, 300dpi, 0.00x0.00 cm, bb=0 0 694 575
  \includegraphics[width=70mm,bb=0 0 694 575,keepaspectratio=true]{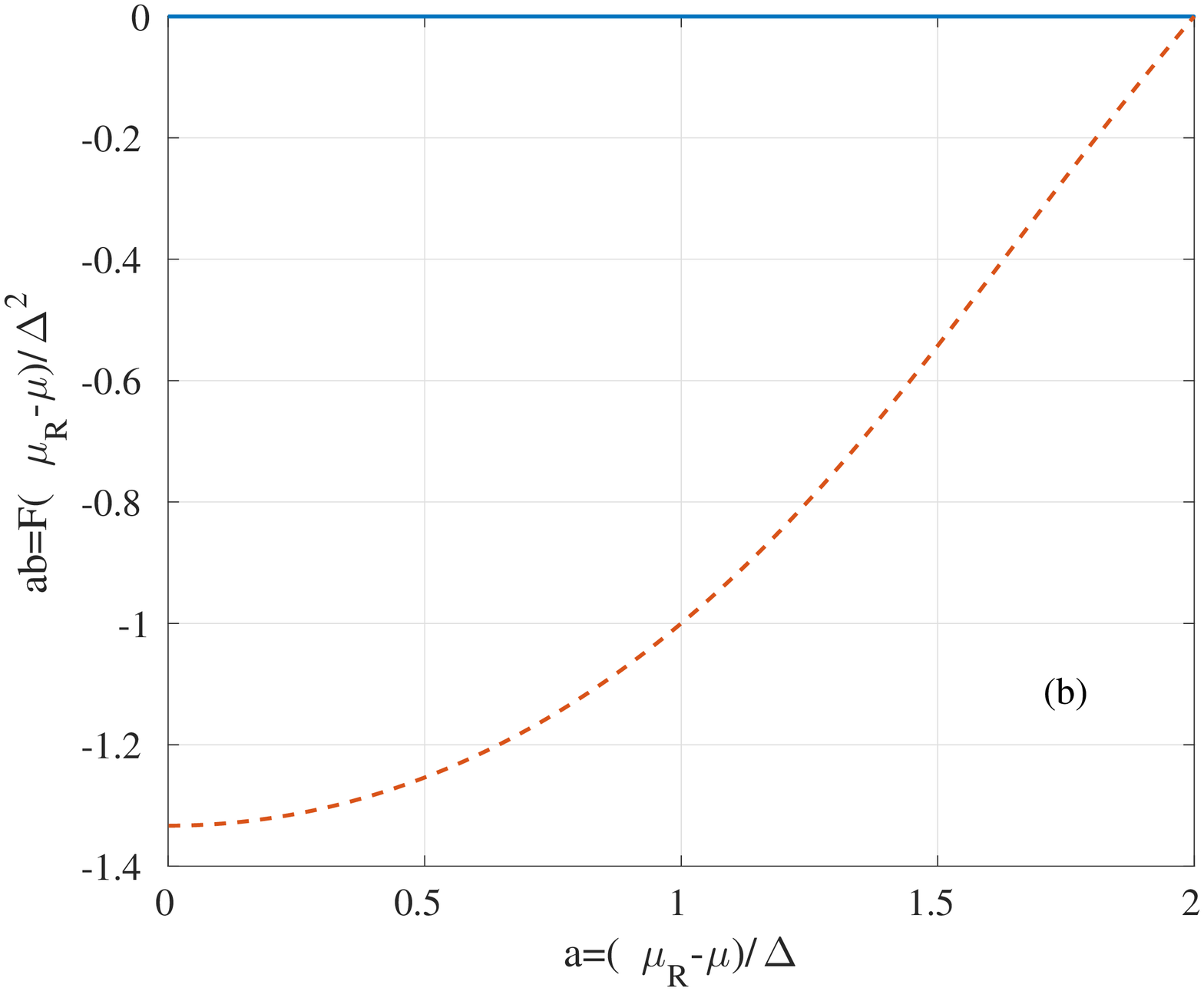}
  \caption{(Color online) In (a) are the solutions $b$ vs $a$, obtained from Eqs. (\ref{Eqs_Delta_xF_T0}) and (\ref{Eqs_Delta_xF_T0_v2}). There are two solutions, $b \equiv 0$ (solid blue line) and $b < 0$ (dashed red line). In (b) we plot the product $ab$ vs $a$. For the negative function (dashed red line), $\lim_{a\searrow 0} ab = 4/3$.}
  \label{b_solution_T0_v2}
\end{figure}

The solutions $b(a)$, from Eqs. (\ref{Eqs_Delta_xF_T0}) and (\ref{Eqs_Delta_xF_T0_v2}), are plotted in Fig. \ref{b_solution_T0_v2}.
We see that if $a < 1$, then $ab < -1$ (see Fig. \ref{b_solution_T0_v2}~b).
In this case, $n_{\xi_x}(T=0) = 1$, for $r = x/y \in [1, r_2)$, and $n_{-\xi_x}(T=0) = 1$, for $r = x/y \in [1, r_0)$.
In the limit $a \searrow 0$, the product $ab$ converges to a constant, which can be readily calculated from Eq. (\ref{Eqs_xF_T0_v2}), namely
\begin{equation}
  \lim_{a\to 0} ab = - 4/3 . \label{lim_ab}
\end{equation}
For $a \in (1,2)$, we have $ab \in (-1, 0)$ and $n_{\xi_x}(T=0) = 1$, if $r = x/y \in (r_0, r_2)$, whereas $n_{-\xi_x}(T=0) = 0$ for any $r$.

\begin{figure}[t]
  \centering
  \includegraphics[width=70mm,bb=0 0 694 554,keepaspectratio=true]{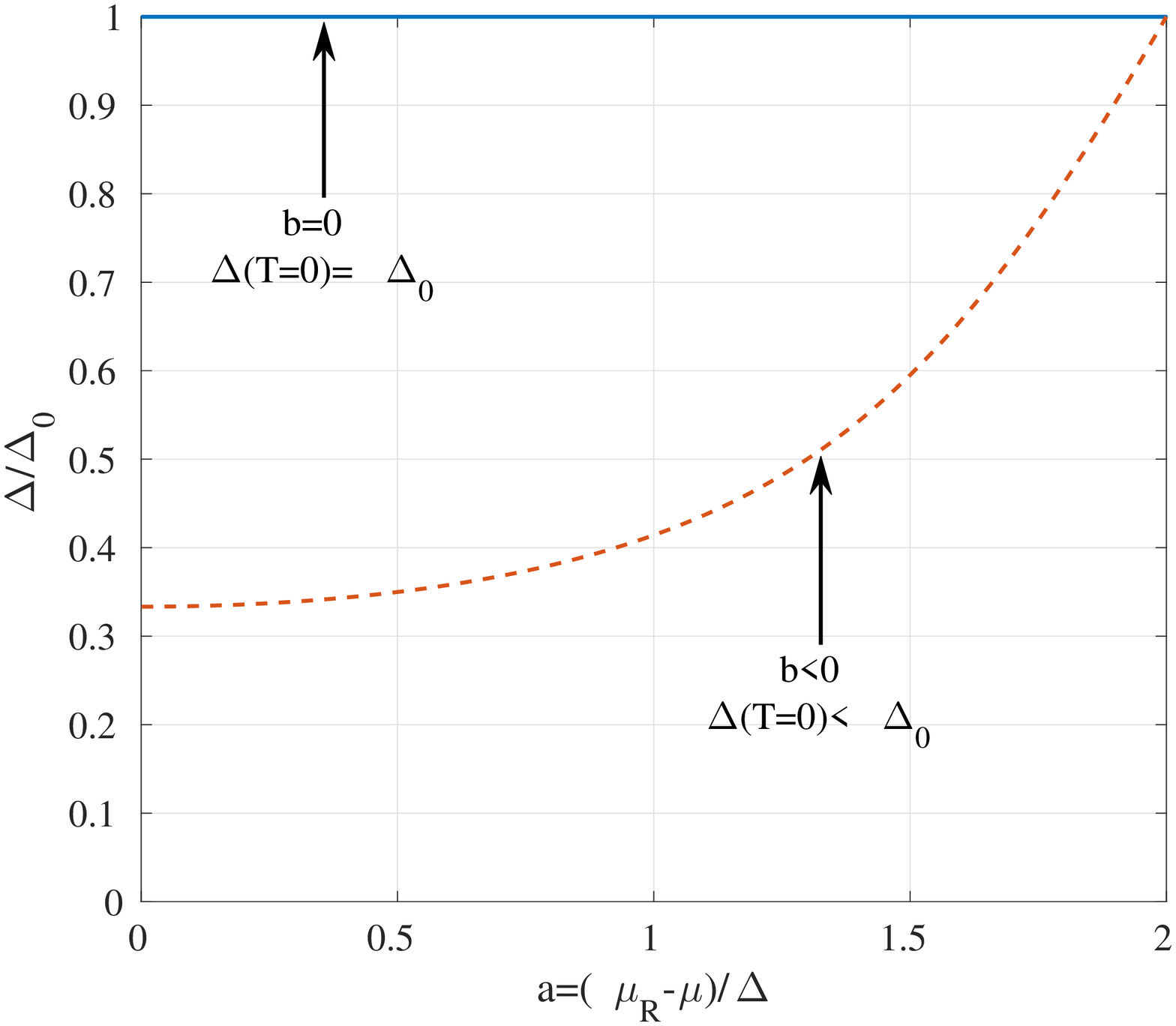}
  % Delta_ratio_T0_v2.eps: 0x0 pixel, 300dpi, 0.00x0.00 cm, bb=0 0 694 554
  \caption{The ratio $\Delta(T=0)/\Delta_0$ vs $a$, for the solutions of $b$ plotted in Fig. \ref{b_solution_T0_v2}.}
  \label{Delta_ratio_T0}
\end{figure}

In Fig.~\ref{Delta_ratio_T0} we plot $\Delta(T=0)/\Delta_0$ for the two solutions of $b$ plotted in Fig. \ref{b_solution_T0_v2}~(a). For the solution $b \equiv 0$, $\Delta(T=0) = \Delta_0$ for any $x$, whereas for the solution $b < 0$, $\Delta(T=0) \le \Delta_0$. Using Eqs. (\ref{Eqs_Delta_T0_1_v2}) and (\ref{lim_ab}) we obtain
\begin{equation}
  \lim_{a\to 0} \Delta^{(b<0)}(T=0) = \Delta_0 /3 , \label{lim_Delta}
\end{equation}
for the solution with $b<0$.

\begin{figure}[t]
  \centering
  \includegraphics[width=7cm,bb=0 0 694 553,keepaspectratio=true]{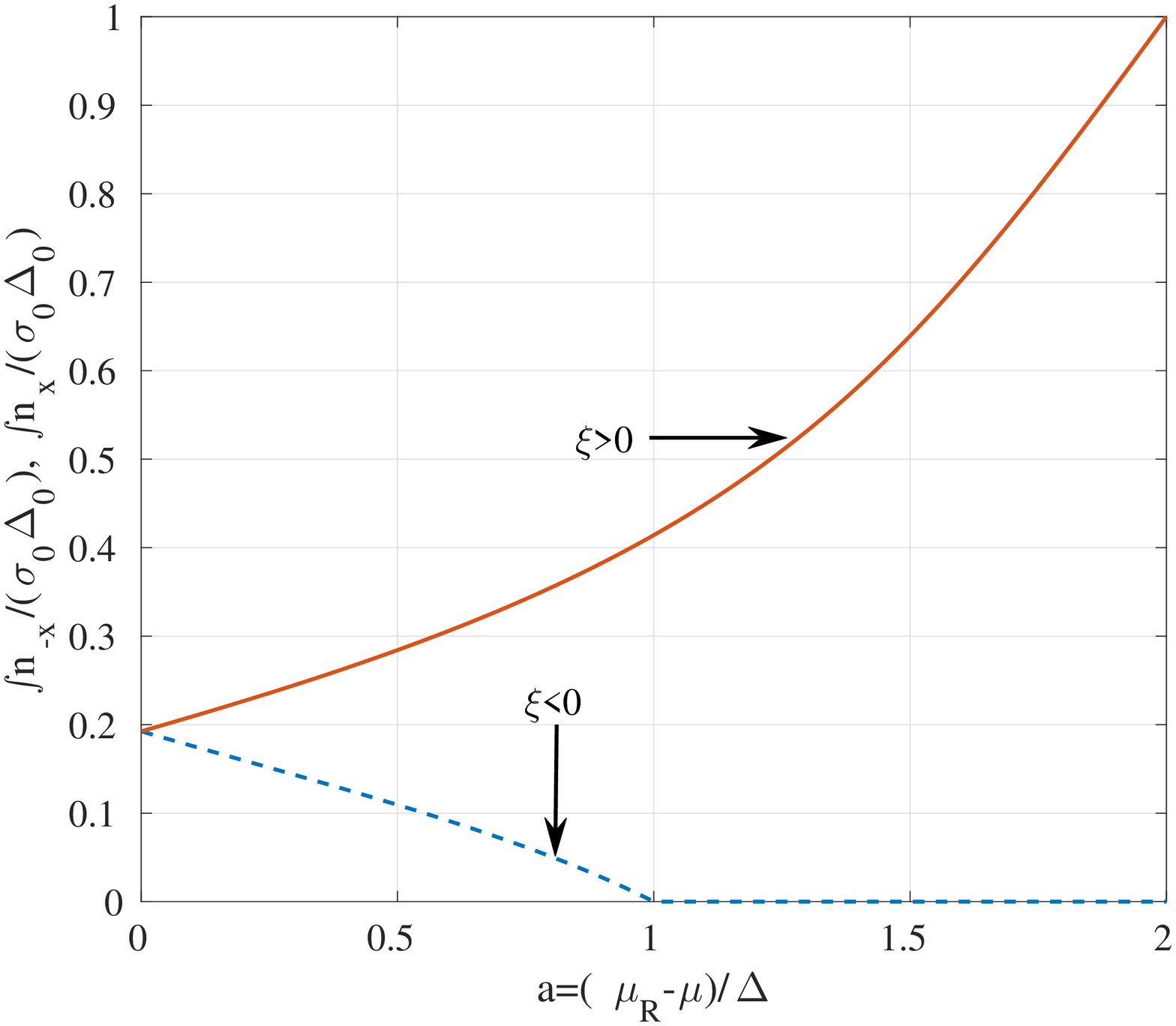}
  % populations.eps: 0x0 pixel, 300dpi, 0.00x0.00 cm, bb=0 0 694 553
  \caption{(Color online) The total populations (integral over the $\xi$ or $\epsilon$, symbolically represented as a primitive) of the branches with $\xi < 0$ (blue dashed line) and $\xi > 0$ (red solid line), for the solutions with $b<0$.}
  \label{populations}
\end{figure}

Having the solutions for $b$ and $\Delta$, we can calculate the quasiparticle populations and the quasiparticle imbalance.
For the solutions with $b=0$, the situation is trivial: $n_{\xi_x} = n_{-\xi_x} = 0$ for any $r$.
For $b<0$, if $1 < a < 2$, then $-1 < ab < 0$ and only the branch with $\xi > 0$ is populated for $\xi \in \left[\Delta\sqrt{r_0^2 - 1}, \Delta\sqrt{r_2^2 - 1}\right]$.
If $0 < a < 1$, then $-4/3 < ab < -1$ and both branches are populated: the branch $\xi < 0$ is populated in the interval $\xi \in \left[- \Delta\sqrt{r_0^2 - 1}, 0\right]$, whereas the branch $\xi > 0$ is populated in the interval $\xi \in \left[0, \Delta\sqrt{r_2^2 - 1}\right]$.
I plot the branches populations in Fig.~\ref{populations} and we observe that the branch imbalance is non-zero for any $a > 0$, whereas $\lim_{a\nearrow 2} n_{\xi_x}/(\sigma_0\Delta_0) = 1$.
When $a \searrow 0$, the population imbalance disappears, although $b\to-\infty$ and $\Delta \searrow \Delta_0/3$.
This situation corresponds to $\mu_R = \mu$, $\Delta = \Delta_0/3$, and the population for each branch equal to $\sigma_0\Delta_0/(3\sqrt{3})$.

\begin{figure}[t]
  \centering
  \includegraphics[width=7 cm,bb=0 0 694 555,keepaspectratio=true]{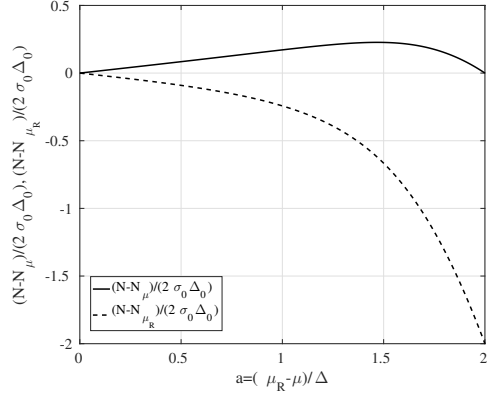}
  % NNmuR_tot.eps: 0x0 pixel, 300dpi, 0.00x0.00 cm, bb=0 0 694 555
  \caption{The difference between the total number of particles $N$ and number of free-particle states up to $\mu$ ($N_\mu$) and $\mu_R$ ($N_{\mu_R}$) vs $a = (\mu_R - \mu)/\Delta$.}
  \label{NNmuR_tot}
\end{figure}

The total number of particles is \cite{PhysicaA.464.74.2016.Anghel}
\begin{eqnarray}
  N &\equiv& \langle \hat N \rangle = N' + \sum_{\bk}2v^2_{\bk} + \sum_{\bk, i} n_\bk \frac{\xi_\bk}{\epsilon_\bk}
  , \label{N_exp_val1}
\end{eqnarray}
where $N'$ is the contribution coming from outside the attraction band.
Introducing the notation $N_\mu \equiv 2 \sum_{\bk}^{k\le k_F} 1$ we get
%
% \begin{subequations} \label{defs_Ntot}
\begin{eqnarray}
  N &=& N_\mu - \sum_\bk^{-\hbar\omega_c\le\xi_\bk<0} \left[ 1 - \frac{|\xi_\bk|}{\epsilon_\bk} + \left( n_{\bk 0} + n_{\bk 1} \right) \frac{|\xi_\bk|}{\epsilon_\bk} \right] \nonumber \\
  && + \sum_\bk^{0\le\xi_\bk \le \hbar\omega_c} \left[ 1 - \frac{\xi_\bk}{\epsilon_\bk} + \left( n_{\bk 0} + n_{\bk 1} \right) \frac{\xi_\bk}{\epsilon_\bk} \right] , \label{N_Nmu_sum}
\end{eqnarray}
which, in the quasicontinuous limit (assuming constant DOS) becomes
\begin{eqnarray}
  N % &=& N_\mu + \sigma_0 \int_{-\hbar\omega_c}^{\hbar\omega_c} ( n_{\xi 0} + n_{\xi 1}) \frac{\xi}{\epsilon} d\xi \nonumber \\
  &=& N_\mu + 2 \sigma_0 \int_{-\hbar\omega_c}^{\hbar\omega_c} \frac{\xi n_\xi}{\epsilon_\xi} d\xi
%   &=& N_\mu + 2 \sigma_0 \int_{\Delta}^{\hbar\omega_c} ( n_{\sqrt{\epsilon^2 - \Delta^2}} - n_{-\sqrt{\epsilon^2 - \Delta^2}}) d\epsilon
  \label{N_Nmu_int_sconst2}
\end{eqnarray}
% \end{subequations}
%
For the situation when $b\equiv0$ and $\Delta = \Delta_0$, $n_{\xi_x} = n_{-\xi_x} = 0$ and $N = N_\mu$ for any $a$.
In the case of solutions with $b<0$, from Eq. (\ref{N_Nmu_int_sconst2}) we obtain
\begin{subequations} \label{NN_tot_T0}
\begin{equation}
  \frac{N -N_\mu}{2\sigma_0} % = \int\limits_{-\Delta \sqrt{r_0^2-1}}^{\Delta \sqrt{r_2^2-1}} \frac{\xi}{\epsilon} d \xi
  = \int\limits_{\Delta \sqrt{r_0^2-1}}^{\Delta \sqrt{r_2^2-1}} \frac{\xi \, d \xi}{\sqrt{\xi^2 + \Delta^2}}
  % = \Delta \int\limits_{\sqrt{r_0^2-1}}^{\sqrt{r_2^2-1}} \frac{z \, d z}{\sqrt{z^2 + 1}}
  = \Delta (r_2-r_0) . \label{N_tot_T0}
\end{equation}
If we denote by $N_{\mu_R} \equiv 2 \sum_\bk^{\epsilon^{(0)}\le\mu_R}$ the number of free-particle states up to $\mu_R$, then
\begin{equation}
  \frac{N -N_{\mu_R}}{2\sigma_0} = \frac{N -N_\mu}{2\sigma_0} - (\mu_R - \mu)
  = \Delta (r_2-r_0 - a) . \label{N_tot_T0R}
\end{equation}
\end{subequations}
%
% Both, $N -N_\mu$ and $N -N_{\mu_R}$, are plotted in Fig. \ref{N_tot}.
For $a\in (0,2)$, $N -N_\mu > 0$, whereas $N -N_{\mu_R} < 0$.
For $a=0$, $N -N_{\mu} = N-N_{\mu_R} = 0$, whereas for $a=2$, $N -N_{\mu} = 0$ and $N -N_{\mu_R} = -4\sigma_0\Delta_0$ (see Fig.~\ref{NNmuR_tot}).

Let us now analyze the partition function
\begin{eqnarray}
  \ln(\cZ)_{\beta\mu} &=& - \sum_{\bk i} [(1 - n_{\bk i}) \ln(1 - n_{\bk i}) + n_{\bk i} \ln n_{\bk i} ] \nonumber \\
  && - \beta (E-\mu_R N) , \label{cZ_beta_mu}
\end{eqnarray}
and denote $\lim_{T\to0} \cZ \equiv \cZ_0$. We observe that 
\begin{eqnarray}
  && - \frac{ k_BT \ln \cZ_0 }{2\sigma_0} = \frac{ E - \mu_R N }{2\sigma_0} % = \int_{-\hbar\omega_c}^{\hbar\omega_c} \left[ \epsilon_\xi - (\mu_R - \mu) \frac{\xi}{\epsilon_\xi} \right] \nonumber \\
  %
  % && \times n_\xi \, d\xi 
  = \frac{\cE_0}{2\sigma_0} + \int_0^{\hbar\omega_c} \left[ \epsilon_\xi \left(n_\xi + n_{-\xi}\right) \right. \nonumber \\
  && \left. - \left(\mu_R - \mu\right) \frac{\xi}{\epsilon} \left(n_\xi - n_{-\xi}\right) \right] d\xi
  \equiv \frac{\cE_0}{2\sigma_0} + I_1 + I_2 , \label{logZ_T0}
\end{eqnarray}
where \cite{PhysicaA.464.74.2016.Anghel}
\begin{eqnarray}
  \cE_0 &=& - \frac{\sigma_0 \Delta^2}{2} \left[ 1 + 2 \ln \left( \frac{\Delta_0}{\Delta} \right) \right] . \label{def_cE0_int}
\end{eqnarray}
The first and second integrals in (\ref{logZ_T0}) are
\begin{subequations} \label{defs_I12}
\begin{eqnarray}
  I_1 % &=& \int_\Delta^{\hbar\omega_c} \frac{ \epsilon^2 }{\sqrt{\epsilon^2 - \Delta^2}} \left(n_{\xi_\epsilon} + n_{-\xi_\epsilon}\right) d\epsilon \nonumber \\
  &=& \Delta^2 \int_1^{\hbar\omega_c/\Delta} \frac{ r^2 }{\sqrt{r^2 - 1}} \left(n_{\xi_x} + n_{-\xi_x}\right) \, dr \label{def_I1}
\end{eqnarray}
and
\begin{eqnarray}
  I_2 % &=& - \left(\mu_R - \mu\right) \int_\Delta^{\hbar\omega_c} \left(n_{\xi_\epsilon} - n_{-\xi_\epsilon}\right) d\epsilon \nonumber \\
  &=& - \Delta^2 a \int_1^{\hbar\omega_c/\Delta} \left(n_{\xi_x} - n_{-\xi_x}\right) \, dr , \label{def_I2}
\end{eqnarray}
\end{subequations}
respectively, where, as before, $x = \beta\epsilon = r y$.

\begin{figure}[t]
  \centering
  \includegraphics[width=7 cm,bb=0 0 695 587,keepaspectratio=true]{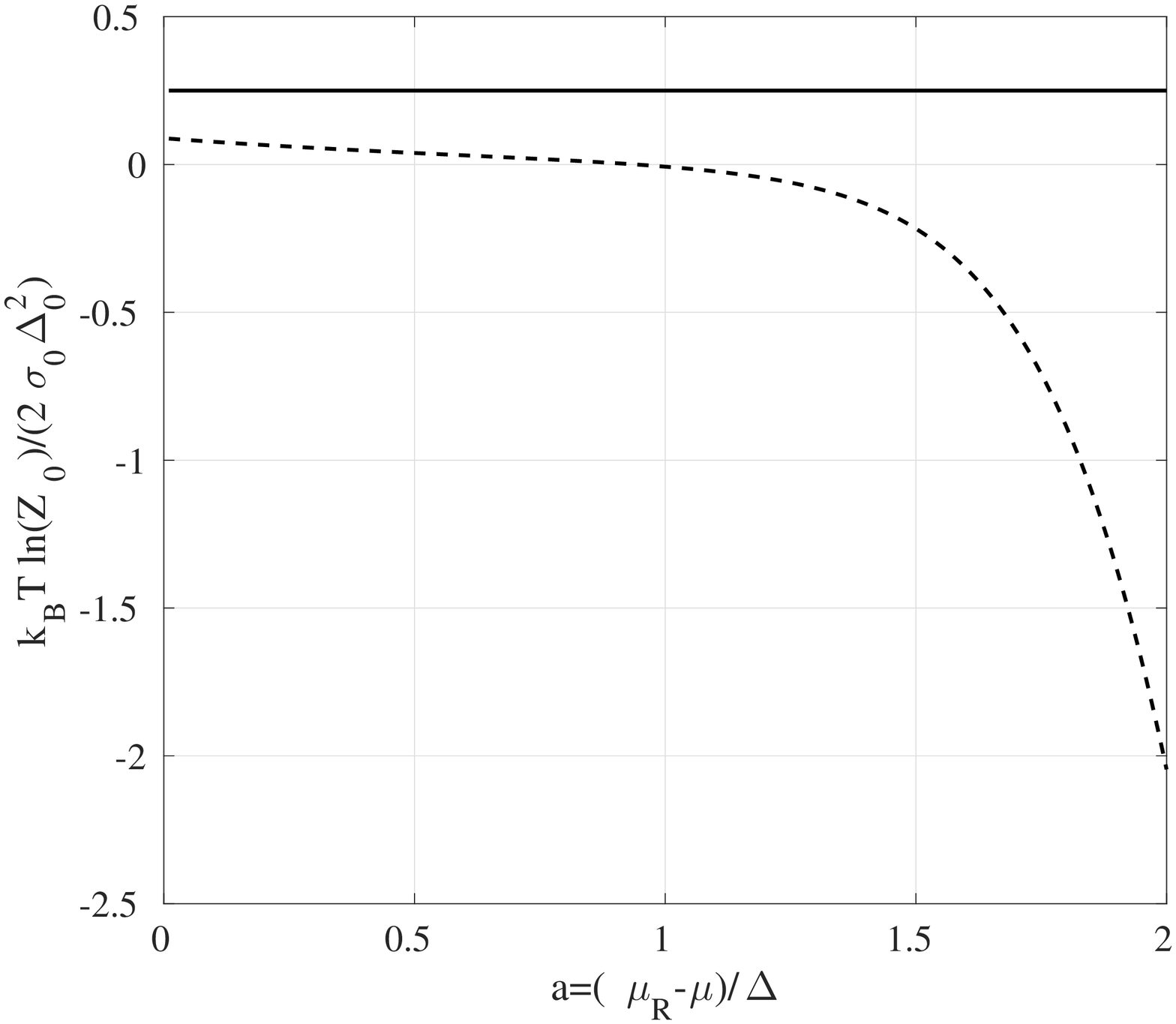}
  % lnZ_T0.eps: 0x0 pixel, 300dpi, 0.00x0.00 cm, bb=0 0 695 587
  \caption{The logarithm of the partition function in the limit $T\to 0$ for the solutions with $b=0$ (solid line) and $b<0$ (dashed line).}
  \label{lnZ_T0}
\end{figure}

For the solution with $b=0$, $n_\xi = n_{-\xi} = 0$ for any $\xi$, so
\begin{equation}
  - \frac{ k_BT \ln \cZ_0^{(b=0)} }{2\sigma_0} = \frac{\cE_0}{2\sigma_0} = - \frac{\Delta_0^2}{4} . \label{logZ_T0_b0}
\end{equation}
For the solution with $b<0$, if $a\le 1$, then $ab \le -1$ and the integral $I_1$ becomes
\begin{subequations} \label{defs_I12_a1}
\begin{eqnarray}
  I_1 &=& \frac{\Delta^2}{2} \left[ r_2 \sqrt{r_2^2-1} + r_0 \sqrt{r_0^2-1} + \ln\left(r_2 + \sqrt{r_2^2-1}\right) \right. \nonumber \\
  && \left. + \ln\left(r_0 + \sqrt{r_0^2-1} \right) \right] , \label{def_I1_a1}
  %
%   I_2 &=& - \Delta^2 a \left( r_2 - r_0 \right) . \label{def_I2_a1}
\end{eqnarray}
% \end{subequations}
%
whereas if $1 < a < 2$, then $ab > -1$ and
%
% \begin{subequations} \label{defs_I12_a2}
\begin{eqnarray}
  I_1 &=& \frac{\Delta^2}{2} \left[ r_2 \sqrt{r_2^2-1} - r_0 \sqrt{r_0^2-1} + \ln\left(r_2 + \sqrt{r_2^2-1}\right) \right. \nonumber \\
  && \left. - \ln\left(r_0 + \sqrt{r_0^2-1} \right) \right] . \label{def_I1_a2}
\end{eqnarray}
The integral $I_2$ has the same expression for any $a\in[0,2]$, namely
\begin{eqnarray}
  I_2 &=& - \Delta^2 a \left( r_2 - r_0 \right) . \label{def_I2_a2}
\end{eqnarray}
\end{subequations}
Pugging Eqs. (\ref{defs_I12_a1}) into (\ref{logZ_T0}) we can calculate the logarithm of the partition function, which is plotted in Fig.~\ref{lnZ_T0}.
We observe that the solutions with $b=0$ are the stable ones, since the partition function takes the bigger value.
The solutions with $b<0$ still satisfy the extremum condition and therefore represent metastable solutions.

\section{Conservation of the number of particles} \label{sec_Nconst}

We see from Fig.~\ref{NNmuR_tot} that $N$ is in general different from both $N_\mu$ and $N_{\mu_R}$.
Since the DOS is constant, in the normal metal phase, the number of particles is $N_{\mu_R}$. Therefore, upon condensation, the number of particles decreases for the situations depicted in Fig.~\ref{NNmuR_tot}.
If we take into account the charge of the electron, this would lead to charging effects and therefore would not be physical.
To solve this problem, we must work in the canonical ensemble, i.e. we must impose that the number of particles is the same before and after the condensation.
If $\mu = \mu_R$, the problem is very simple, since $N_\mu = N_{\mu_R} \equiv N$ for both solutions of the system~(\ref{def_xF_sigma0_set}).
On the other hand, if $\mu_R > \mu$, then for the solution with $b \equiv 0$ and $\Delta = \Delta_0$, the number of particles in the normal phase is $N_{\mu_R}$, whereas in the superconducting phase is $N_\mu (\ne N_{\mu_R})$. Therefore the solutions with $b = 0$ cannot conserve the number of particles when $\mu_R \ne \mu$, so they are not the physical solutions.
The only physical solutions left for $N \ne N_\mu$ are the ones with $b<0$, which are analyzed below.

\begin{figure}[t]
  \centering
  \includegraphics[width=7cm,bb=0 0 695 593,keepaspectratio=true]{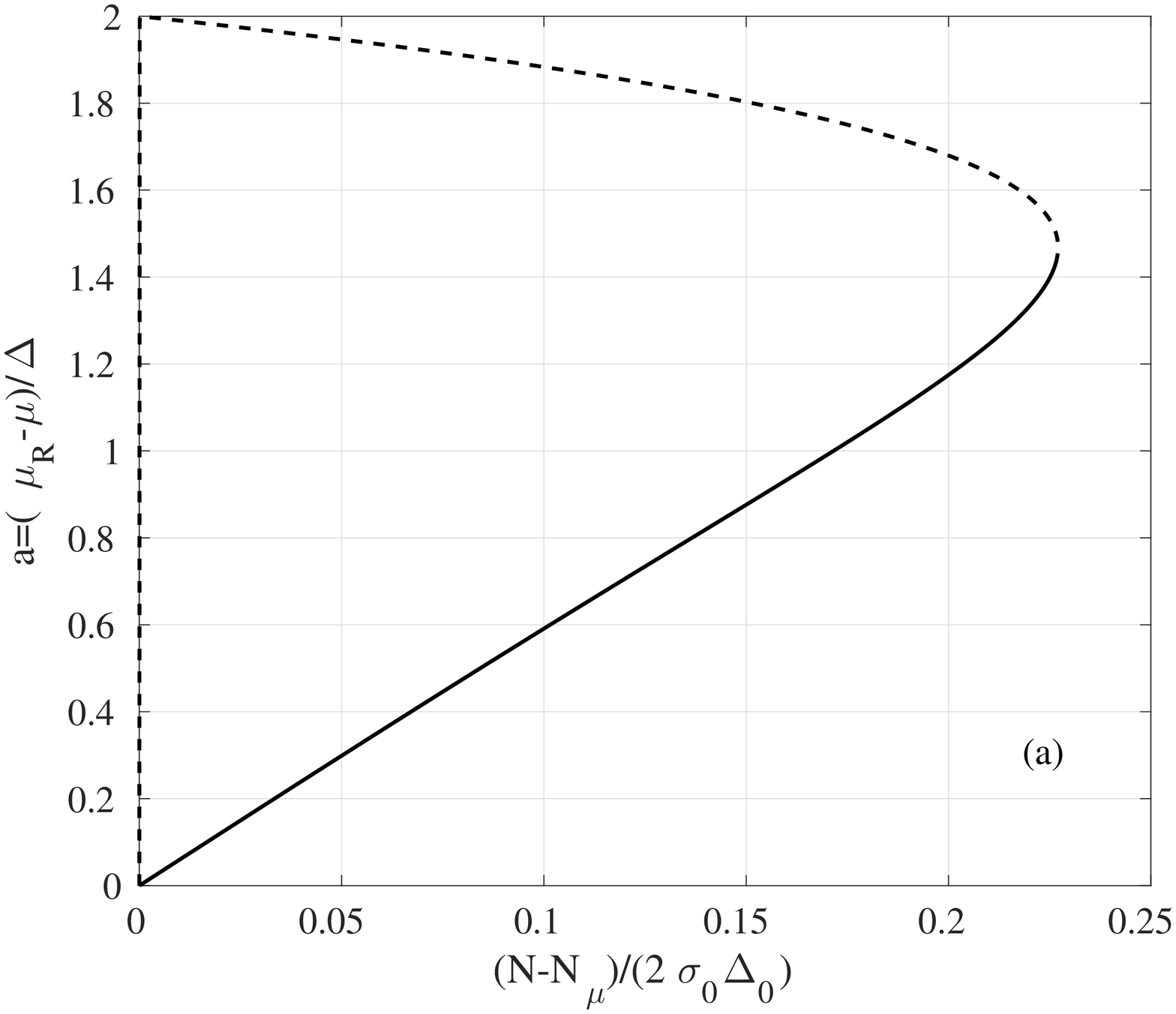} \\
  \includegraphics[width=7cm,bb=0 0 695 593,keepaspectratio=true]{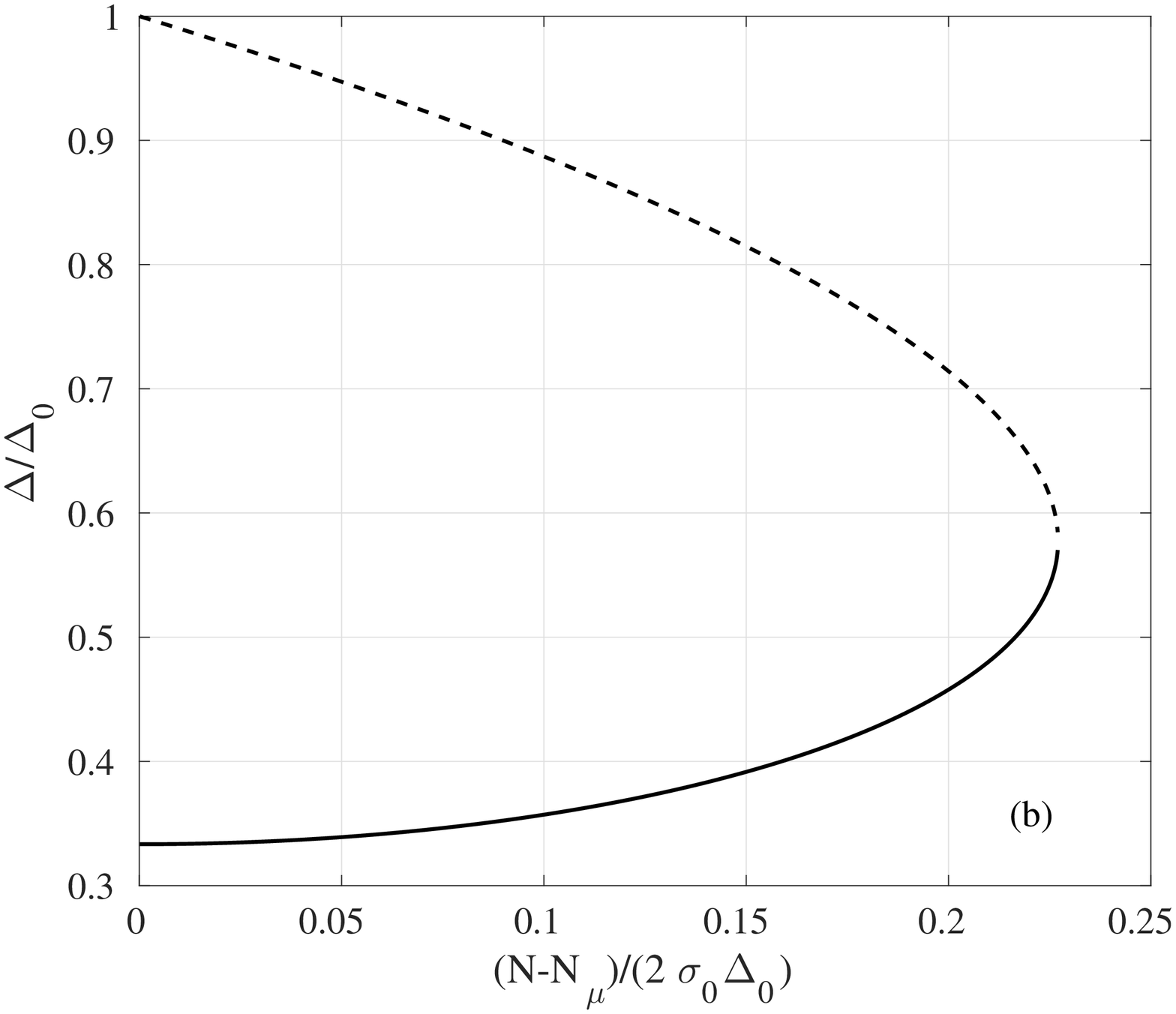}
  % a_Nconst.eps: 0x0 pixel, 300dpi, 0.00x0.00 cm, bb=0 0 695 593
  \caption{The conservation of the number of particles. For each $N$ between $N_\mu$ and $N_{max}$ we can find two values of $\mu_R$ (a), such that $N = N_{\mu_R}$ in the superconducting phase. In (b) we plot the values of the energy gap corresponding to the $a$ values from (a).}
  \label{a_Nconst}
\end{figure}

We can see from Fig.~\ref{NNmuR_tot} that for $\mu$ (and $N_\mu$) fixed, $N \le N_{max} \approx 0.227 (2\sigma_0\Delta_0) + N_\mu \equiv N(a_{max},\mu)$, where $N_{max}$ is the maximum value reached by $N$, when $a = a_{max} \approx 1.468$.
Therefore, if the number of particles satisfies the relations $N_\mu < N < N_{max}$, then the superconducting phase may be reached for two values of $a$, such that $N = N_{\mu_R}$.
These values are denoted by $a_{1,2}$, where $0 < a_1 < a_{max}$ and $a_{max} < a_2 < 2$.
If $N = N_{max}$, then $a_1 = a_2 = a_{max}$, whereas if $N = N_\mu$, then $a = 0$, but we have solutions for both, $b = 0$ and $b \to -\infty$ (the second solution corresponds to $\Delta = \Delta_0/3$).
These solutions are plotted in Fig.~\ref{a_Nconst}(a), whereas the values of the energy gap corresponding to them are plotted in Fig.~\ref{a_Nconst}(b).

%%%%%%%%%%%%%%%%%%%%%%%%%%%%%%%%%%%%%%%
\section{Conclusions} \label{discussion}

I analyzed the BCS formalism in the low temperature limit, under the assumption that the attraction band is asymmetric with respect to the chemical potential of the system $\mu_R$.
I denoted by $\mu$ the center of the attraction band and by $\Delta_0$ the energy gap in the standard BCS theory, at zero temperature.
First, I analyzed the system under the assumption that $\mu_R$ is the same in both, superconducting phase and normal phase, and then by imposing that the total number of particles is the same in both phases.

In the first case I observed that if $|\mu_R - \mu| > 2\Delta_0$, the system of equations (\ref{def_xF_sigma0_set}), which gives the energy gap, has no solutions, so the superconducting state cannot exist even at zero temperature.
If $|\mu_R - \mu|/\Delta_0 \in (0,2)$, the system (\ref{def_xF_sigma0_set}) has two solutions: one with $\Delta(T=0) = \Delta_0$ and another one, with $\Delta(T=0) < \Delta_0$ (see Fig. \ref{Delta_ratio_T0}) and a quasiparticle imbalance which appears in equilibrium (see Fig. \ref{populations}).
The solutions with $\Delta(T=0) = \Delta_0$ are stable, whereas the other ones are metastable, since in the first case the partition function takes bigger values.

It is interesting to note that in the limit $\mu_R \to \mu$, when the standard BCS theory should be obtained, the system (\ref{def_xF_sigma0_set}) still has two solutions, one with $\Delta(T=0) = \Delta_0$ (as expected) and another one, with $\Delta(T=0) = \Delta_0/3$.
In both of these solutions the quasiparticle imbalance disappears.

Regarding the second case analyzed, the change of the number of particles, when going from the normal metal state to the superconducting state (Eqs.~\ref{NN_tot_T0}), would lead to charging effects if the Coulomb interaction is taken into account.
Therefore, in Section~\ref{sec_Nconst} I imposed the conservation of the number of particles.
In this case, the chemical potential of the reservoir, $\mu_R$, is determined by the total number of particles and the condition $N \equiv N_{\mu_R}$ in the normal metal state, where $N_{\mu_R}$ is the number of states below $\mu_R$.
If $N_\mu$ is the number of states below $\mu$, then if $\mu = \mu_R$, then $N = N_\mu = N_{\mu_R}$ and we recover the BCS solution, as stated above.
If $N > N_\mu$, there is a maximum value $N_{max} \approx 0.227 (2\sigma_0\Delta_0) + N_\mu$, up to which solutions to the problem exist.
For each $N$ in between $N_\mu$ and $N_{max}$, we can find two values of $\mu_R$ which satisfy the conservation of $N$.

It is interesting to note that if $N > N_\mu$, one cannot find a solution for $b = 0$ and $\Delta = \Delta_0$, but only solutions with $b < 0$ and $\Delta < \Delta_0$. This means that for $N \ne N_\mu$, the stable solutions are unphysical and only the ``metastable'' ones conserve the number of particles.

The formalism is symmetric under the simultaneous exchanges $\mu_R - \mu \to - (\mu_R - \mu)$, $x_F \to -x_F$, and $\xi \to -\xi$.
Therefore the results presented can be easily extended to $\mu_R < \mu$ (i.e. $a < 0$) and $N < N_\mu$.

\section{Acknowledgments}

Discussions with Dr. G. A. Nemnes are gratefully acknowledged.
This work has been financially supported by CNCSIS-UEFISCDI (project IDEI 114/2011) and ANCS (project PN-09370102 PN 16420101/ 2016). Travel support from Romania-JINR Collaboration grants 4436-3-2015/2017, 4342-3-2014/2015, and the Titeica-Markov program is gratefully acknowledged.


\begin{thebibliography}{10}

\bibitem{PhysicaA.464.74.2016.Anghel}
D.~V. Anghel and G.~A. Nemnes.
\newblock {\em Physica A}, 464:74, 2016.

\bibitem{PhysRev.108.1175.1957.Bardeen}
J.~Bardeen, L.~N. Cooper, and J.~R. Schrieffer.
\newblock {\em Phys. Rev.}, 108:1175, 1957.

\bibitem{Tinkham:book}
Michael Tinkham.
\newblock {\em Introduction to Superconductivity}.
\newblock McGraw Hill, Inc., 2 edition, 1996.

\bibitem{NuclPhysA.887.1.2012.Parvan}
A.S. Parvan.
\newblock {\em Nucl. Phys. A}, 887:1, 2012.

\bibitem{LowTempPhys.41.112.2015.Parfeniev}
R.~V. Parfeniev, V.~I. Kozub, G.~O. Andrianov, D.~V. Shamshur, A.~V. Chernyaev,
  N.~Yu. Mikhailin, and S.~A. Nemov.
\newblock {\em Low Temp. Phys.}, 41:112, 2015.

\bibitem{PhysRevLett.87.047001.2001.Bouquet}
F.~Bouquet, R.~A. Fisher, N.~E. Phillips, D.~G. Hinks, and J.~D. Jorgensen.
\newblock {\em Phys. Rev. Lett.}, 87:047001, 2001.

\bibitem{PhysRevLett.87.137005.2001.Szabo}
P.~Szab\'o, P.~Samuely, J.~Ka\ifmmode \check{c}\else
  \v{c}\fi{}mar\ifmmode~\check{c}\else \v{c}\fi{}\'{\i}k, T.~Klein, J.~Marcus,
  D.~Fruchart, S.~Miraglia, C.~Marcenat, and A.~G.~M. Jansen.
\newblock {\em Phys. Rev. Lett.}, 87:137005, 2001.

\bibitem{Science.314.1910.2006.Tanaka}
K.~Tanaka, W.~S. Lee, D.~H. Lu, A.~Fujimori, T.~Fujii, Risdiana, I.~Terasaki,
  D.~J. Scalapino, T.~P. Devereaux, Z.~Hussain, and Z.-X. Shen.
\newblock {\em Science}, 314(5807):1910, 2006.

\bibitem{PhysRevLett.98.267004.2007.Kondo}
T.i Kondo, T.~Takeuchi, A.~Kaminski, S.~Tsuda, and S.~Shin.
\newblock {\em Phys. Rev. Lett.}, 98:267004, 2007.

\bibitem{PhysRevLett.28.1363.1972.Clarke}
J.~Clarke.
\newblock {\em Phys. Rev. Lett.}, 28:1363, 1972.

\bibitem{PhysRevLett.28.1366.1972.Tinkham}
M.~Tinkham and J.~Clarke.
\newblock {\em Phys. Rev. Lett.}, 28:1366, 1972.

\bibitem{PhysRevB.6.1747.1972.Tinkham}
M.~Tinkham.
\newblock {\em Phys. Rev. B}, 6:1747, 1972.

\bibitem{PhysRevB.22.4346.1980.Smith}
A.~D. Smith, M.~Tinkham, and W.~J. Skocpol.
\newblock {\em Phys. Rev. B}, 22:4346, 1980.

\bibitem{PhysRevB.21.3879.1980.Smith}
A.~D. Smith, W.~J. Skocpol, and M.~Tinkham.
\newblock {\em Phys. Rev. B}, 21:3879, 1980.

\bibitem{PhysRevLett.72.558.1994.Hirsch}
J.~E. Hirsch.
\newblock {\em Phys. Rev. Lett.}, 72:558, 1994.

\bibitem{PhysRevB.58.8727.1998.Hirsch}
J.~E. Hirsch.
\newblock {\em Phys. Rev. B}, 58:8727, 1998.

\bibitem{PhysScr.88.035704.2013.Hirsch}
J.~E. Hirsch.
\newblock {\em Phys. Scr.}, 88:035704, 2013.

\bibitem{PhysRevB.39.11515.1989.Hirsch}
J.~E. Hirsch and F.~Marsiglio.
\newblock {\em Phys. Rev. B}, 39:11515, 1989.

\bibitem{PhysRevB.41.6435.1990.Marsiglio}
F.~Marsiglio and J.~E. Hirsch.
\newblock {\em Phys. Rev. B}, 41:6435, 1990.

\end{thebibliography}
\end{document}